\documentclass[preprint,aps,showpacs]{revtex4}
\usepackage{graphicx}% Include figure files
\usepackage{dcolumn}% Align table columns on decimal point
\usepackage{bm}% bold math

\begin{document}
 
\title{Systematic investigation of the 
structure of the Si(553)-Au surface from first principles} 
\author{Sampsa Riikonen}
\email{swbriris@sc.ehu.es}
\affiliation{Departamento de F\'{\i}sica de Materiales,
Facultad de Qu\'{\i}mica, Universidad del Pa\'{\i}s Vasco (UPV/EHU),
Apdo. 1072,
20080 San Sebasti\'an, Spain}
\affiliation{Donostia International Physics Center (DIPC),
Paseo Manuel de Lardizabal 4, 20018 San Sebasti\'an, Spain}
\author{Daniel S\'anchez-Portal}
\email{sqbsapod@sc.ehu.es}
\affiliation{
Centro de F\'{\i}sica de Materiales,
Centro Mixto CSIC-UPV/EHU, Apdo. 1072,
20080 San Sebasti\'an, Spain}
\affiliation{Donostia International Physics Center (DIPC),
Paseo Manuel de Lardizabal 4, 20018 San Sebasti\'an, Spain}

\date{\today}

\begin{abstract}
We present here 
a comprehensive search for the structure
of the Si(553)-Au reconstruction. More than two hundred different
trial structures have been studied using first-principles
density-functional calculations with the SIESTA code. 
An iterative
procedure, with a step-by-step increase
of the  accuracy and computational cost of the calculations, was
used to allow for the study of this large number of configurations.
We have 
considered reconstructions restricted
to the topmost bilayer and studied two types:
{\it i}) ``flat" surface-bilayer models, where
atoms at the topmost bilayer
present different coordinations
and registries with the underlying bulk, and {\it ii}) nine
different models based on the substitution
of a silicon atom by a gold atom
in different positions of a $\pi$-bonded chain reconstruction
of the Si(553) surface. We have developed a compact
notation that allows us to label and identify all these 
structures. This is very useful for the
automatic generation of trial geometries and counting
the number of inequivalent
structures, i.e., structures having different bonding topologies.  
The most stable models
are those that
exhibit a honeycomb-chain structure
at the step edge. One of our models (model f2) reproduces
the main features of the room temperature photoemission and scanning-tunneling
microscopy data. Thus we conclude that 
f2 structure is a good candidate for the high 
temperature
structure of the Si(553)-Au surface.
\end{abstract}
 
\pacs{73.20.At, 71.15.Mb, 79.60.Jv, 68.35.Bs}
 
\maketitle
 
\section{Introduction}

During the last few decades several complex surface 
reconstructions that exhibit one-dimensional
metal wires have been extensively studied. 
The interest in these systems has been mainly driven by the theoretical 
prediction that the behavior of a  one-dimensional metal 
should sensibly deviate from the Fermi liquid picture
due to electron correlation.~\cite{1dbook}
However, the coupling between electrons and lattice
vibrations is also enhanced in one dimension and 
these systems present a strong tendency to suffer structural
transitions that reduce their symmetry 
and destroy their metallicity.
The competition between the effects of electron-electron 
and electron-phonon interactions makes the physics
of one-dimensional systems quite a rich and interesting topic. 
In this context, the quasi-one-dimensional 
reconstructions formed by several metals
on flat and vicinal Si(111) surfaces have been regarded as ideal systems 
to realize a one-dimensional metal. On the one hand, 
the existence of an electronic gap in the substrate 
allows for the existence of purely one-dimensional electronic
states near the Fermi energy associated with the 
metal chains in the surface.
On the other hand, it was believed that
in some cases the rigidity of the substrate could allow for
the observation of characteristic features of the Tomonaga-Luttinger liquid in the
photoemission experiments,~\cite{1dbook,segovia99}
avoiding the appearance
of broken-symmetry phases at sufficiently low temperatures.
Furthermore, ordered arrays of well aligned 
wires can be formed over large areas on these susbtrates, which 
allow for accurate measurements using angle resolved photoemission (ARPES).
In the case of vicinal substrates the average terrace width 
can be controlled by the miscut angle, providing a
route to tune inter-wire distances and interactions.~\cite{crain04}

Some examples of 
quasi-one-dimensional reconstructions
that have attracted much attention in recent years
are the  
Si(111)-(5$\times$2)-Au,~\cite{losio00,erwin03,matsuda03,crain04,riikonen05}
Si(557)-Au,~\cite{losio01,sanchezportal02,robinson02,sanchezportal03,ahn03,sanchezportal04,yeom05,krawiec06,riikonen07}
Si(553)-Au,~\cite{crain03,riikonen05_2,ghose05,crain05,ahn05,riikonen06,crain06,snijders06} 
and Si(111)-(4$\times$2)-In~\cite{yeom99,bunk99,cho01,miwa01,ahn04,park04,kurata05,cho05,gonzalez05,gonzalez06,yeomcomment06,riikonen06_2,ahn07} surfaces.
All these systems show clear one-dimensional features both in ARPES and 
scanning tunnelling microscopy (STM)
experiments. At least in the cases of 
the Si(557)-Au, Si(553)-Au and Si(111)-(4$\times$2)-In surfaces 
some of these one-dimensional 
bands seem to be metallic at room temperature. 
However, these three reconstructions suffer structural distortions
associated with metal-insulator transitions when 
the temperature is lowered.~\cite{ahn03,ahn05,snijders06,yeom99,ahn04,park04}

The photoemission spectrum of the Si(557)-Au 
surface is dominated by two proximal one-dimensional bands
with a parabolic-like dispersion. 
According to the {\it ab initio}
calculations based on the existing structural 
model~\cite{robinson02,sanchezportal03} these two
bands are due to the spin-orbit splitting of 
a band coming from the hybridization of the states of the gold chains
with their silicon neighbors.~\cite{sanchezportal04,barke06}
These bands are half-filled and exhibit, 
as the temperature is lowered,
a metal-insulator transition 
that is accompanied by a periodicity doubling
along the steps 
in the STM images.~\cite{ahn03,sanchezportal04,yeom05,riikonen07}

The Si(553)-Au surface, with a smaller terrace width,
shows two bands similar to those observed
in the case of the Si(557)-Au system. It also shows
a strongly-dispersing one-dimensional band with 
a fractional ($\sim{1\over3}$) filling.~\cite{crain03,crain04}
It was proposed that this 
one-third-filled band could provide the appropriate conditions
to observe large spin-charge separation.
This is in contrast with half-filled
bands, which are unstable against a Mott-Hubbard transition for
large values of the electron-electron interaction, preventing the
observation of a Luttinger metal~\cite{crain04}.
Unfortunately, the Si(553)-Au surface also 
undergoes a metal-insulator transition at low temperatures. 
The Si(553)-Au surface exhibits
two Peierls-like distortions 
with a $\times$2 and $\times$3 increase of the unit cell 
of the chains
along the steps. This is consistent with the occupations
of the bands mentioned above. Furthermore, the $\times$3
periodicity seems to be associated with the step-edge,
while the $\times$2 periodicity appears in the middle
of the terraces. This is consistent with the origin of 
the half-filled bands in the gold wires. {\it Ab initio}
calculations indicate that the gold atoms prefer to occupy
in silicon substitutional positions in the middle of the 
terraces.~\cite{sanchezportal02,sanchezportal03,riikonen05_2}

A good structural model
is necessary to understand in detail the physics of this system.
However, in contrast to the case of the Si(557)-Au surface,
the structure of the Si(553)-Au reconstruction is not well established.
Recently Ghose {\it et al.}~\cite{ghose05} proposed
a structure based on the analysis of x-ray diffraction 
measurements. This model contains a double row of Au atoms
decorating the step edge. This seems inconsistent with
the experimental gold coverage and the model has proven 
to be unstable.~\cite{riikonen06} In a recent work~\cite{riikonen05_2} 
we studied,
using first-principles density-functional calculations, 
five different structural models based on an analogy with 
the structure of the Si(557)-Au surface and an earlier
proposal by Crain {\it et al.}~\cite{crain04} with a
single gold chain.
The most stable structures were those in which the 
step edge is formed
by honeycomb-chain.~\cite{erwin98} 

In the present paper we perform a more 
comprehensive search of the structure
of the Si(553)-Au reconstruction. We have considered
models in which the topmost bilayer contains
eight atoms per 5$\times$1 unit cell, including
one gold atom,  with all 
different coordinations and registries 
with the underlying bulk. We refer to these models as
``flat'' surface-bilayer structures and we have
found that there are 210 inequivalent structures of this
type when reasonable physical constraints are applied.
These structures are generated automatically and 
an iterative
procedure, with a step-by-step increase
of the  accuracy and computational cost of the calculations, is
used to study this large number of configurations.
We have also considered nine different structures
obtained by the substitution of a silicon atom by a gold atom
in different positions of a $\pi$-bonded chain reconstruction
of the Si(553) surface.

As we will see, our most stable structures belong
to the set of flat-bilayer models and exhibit a honeycomb-chain
structure in the step edge. This in agreement with the observation
of our previous study. Indeed the two most stable 
structures (f2 and f4 models) correspond to structures
that have already been explored in Ref.~\onlinecite{riikonen05_2}.
This confirms that the analogy with the Si(557)-Au surface provides
a good route to study the structure of the Si(553)-Au reconstruction and, 
most probably, other related surfaces.~\cite{ahn04_2}
The simulated STM images  of the f2 and f4 models are 
in good agreement with the available 
experimental data.~\cite{snijders06,Ryang07}
Their band structures also reproduce the main features of
the photoemission data.~\cite{crain03,ahn05}
In particular, the band structure of the f2 model 
shows 
two dispersive one-dimensional bands and one of them,
associated with a silicon dangling-bond in the surface, 
has a small fractional occupation in agreement
with the measurements. Therefore, we
conclude that the f2 model is a good candidate 
for the structure of the Si(553)-Au.

Finally, we
have developed a compact
notation that allows us to label and identify all the structures.
This notation is instrumental for the
automatic generation of trial geometries and for counting
the number of inequivalent (having different bonding topologies)
structures.
This notation is applicable to other similar surfaces and
we think that it will be useful for other structural 
investigations.

\section{Methods}
\subsection{Density functional calculations}
\label{sec:comp}
Most of the calculations were performed using
the SIESTA code~\cite{sanchezportal97,soler02}
We used the local density approximation~\cite{pz} and
norm conserving pseudopotentials~\cite{tm}. The gold
pseudopotential included scalar relativistic effects
and was similar to that used in Ref.~\onlinecite{au} and 
in our previous 
calculations~\cite{riikonen05,riikonen05_2,riikonen06,riikonen07}.
Different basis sets were used for the silicon atoms (see below
for more details).
For the fastest calculation a single-$\zeta$ basis (SZ)
of numerical atomic 
orbitals~\cite{artacho99,junquera01,soler02} was used.
The SZ basis uses only
one radial function to represent the orbitals in the 
3$s$ shell and 
another one for the 3$p$ shell.
For more accurate calculations 
double-$\zeta$ basis (DZ), with two different
radial functions for each angular momentum,
and double-$\zeta$ polarized (DZP), including
an additional polarization 3$d$ shell, were used. 
For gold, a DZP basis was always used 
with two different radial functions to represent
the 6$s$ and 5$d$ orbitals and a polarization 6$p$
shell.
The radius of the Si orbitals was
5.25 Bohr for those in the 3$s$ shell and 6.43 Bohr 
for those in the 3$p$ and 3$d$ shells. For
Au the radius of the 6$s$ and 6$p$ orbitals
was 6.24 Bohr and 4.51 Bohr for the orbitals in 
the 5$d$ shell.

We modelled the surface using slabs with different
thicknesses (depending on the desired accuracy).
For the initial and fastest simulations
we used a slab containing only two silicon bilayers,
whereas four bilayers were used to more accurately 
explore the energetics 
of the most favorable structural models.
The atoms in the 
bottom silicon layer were
kept at their ideal bulk positions and 
saturated with hydrogen
atoms. 
A vacuum gap of 15~\AA\ between neighboring slabs was used.
To avoid artificial stresses the lateral lattice parameter was 
always fixed 
to the bulk theoretical value calculated with the similar approximations.  
The lattice parameter depends strongly 
on the basis set used. The calculated values 
are 5.53, 5.48, and 5.42~\AA\ 
for SZ, DZ and DZP basis sets,
respectively.
Therefore, it is necessary to rescale the coordinates 
when the result of a relaxation using a given basis set
is used as the initial guess
for a calculation with a more complete basis.
Systems were relaxed until
the maximum force component along any direction
was less than 0.04 eV/\AA.
Different Monkhorst-Pack~\cite{mp} 
samplings of the surface Brillouin zone were utilized.
They will be referred along this paper as M$\times$N, where
M refers the direction parallel and N perpendicular to the step edge.
A real-space grid equivalent to a 100~Ry plane-wave cut-off was used.
The simulated STM images were
obtained using the Tersoff-Haman theory~\cite{stm}
and the DZP basis set.

For the most stable systems we also performed calculations
using the VASP code \cite{vasp1,vasp2}.
Projected-augmented-wave
potentials and a  well converged
plane-wave basis set with a cutoff of 312~eV were used. All
structures were relaxed (the equilibrium lattice parameter
of bulk silicon obtained with VASP was 5.41~\AA).

\subsection{Strategy of the structural search}
\label{sec:relaxations}

Searching for the lowest-energy configuration of 
any surface reconstruction is a formidable task.
Almost any trial geometry that is not completely 
absurd relaxes to a nearby local
minimum.  One can then generate manually new configurations from 
the most favorable
ones using different physical 
(electron-counting, coordination, etc.) arguments,
hoping to eventually end up in the global minimum.
However, the number of local minima scales
roughly exponentially with the number of atoms involved in 
the reconstruction and the problem becomes intractable
for large cells.
Apart from this heuristic approach one can use more sophisticated
algorithms that
can 
automatically find optimum geometries. Some examples are 
simulated annealing~\cite{kirkpatrick83,cerny85} 
and Monte Carlo simulations
of different types, including
genetic algorithms~\cite{binder86,deaven95,man99}. 
Monte Carlo techniques
have been traditionally used
to find cluster geometries~\cite{bazterra04}, 
and have been recently extended to find 
surface reconstructions~\cite{ciobanu04,ciobanu07}. 
Unfortunately these methods are computationally very expensive
requiring long simulations with thousands of evaluations
of the system energy (and interatomic forces in some cases).
Particularly, genetic algorithms are very powerful but typically 
need hundrends of generations, each one containing tens 
of trial geometries~\cite{bazterra04}. 
Therefore, they are mostly restricted to the use of 
empirical interatomic potentials which, however, 
might not be sufficiently accurate to  reproduce the
energetics of the different geometries explored.

In the present work we adopt a compromise between 
these two ideas. 
Due to the lack of reliable empirical potentials
to represent the interaction between the gold and 
silicon atoms in the surface,
we need 
to explore the energetics of the different
models of the Si(553)-Au reconstruction at the
density functional or similar level of theory.
This precludes the use of Monte Carlo algorithms
to perform a global search of the reconstruction structure.
However, we do not want to restrict our search to explore
a ``few" structural models. Thus, we will rather make a {\it systematic}
search of the optimum surface model within a large family of
physically motivated structures. 
Our approach is the following:

{\it i}) a family of likely structural models for the Si(553)-Au surface 
reconstruction is defined using a heuristic procedure based on 
the analogy with other related and better known surfaces;

{\it ii}) a compact notation is designed to label unambigously each 
of the possible structures within this family;

{\it iii}) from each of these labels, a trial geometry is 
generated automatically and an initial constrained 
relaxation is performed 
to avoid the appearance of unphysical bond distances;

{\it iv}) ``fast" density functional calculations 
using SIESTA are 
used to relax each of the structures
to its closest local energy minimum;

{\it v}) the most stable configurations
from step {\it iv}) are studied using 
more time consuming ``accurate" SIESTA 
calculations;

{\it vi}) finally, since the energy differences between different 
structural models are quite small, we check
the energy ordering of our most stable configurations using 
a different methodology: plane-wave calculations using 
the VASP code.

By ``fast" SIESTA calculations we mean here calculations performed
using small basis sets (i.e, with a small number of basis orbitals
per atom, like SZ or DZ basis sets \cite{sanchezportal97,soler02}), 
limiting the number of k-points and 
optimized degrees of freedom and/or using 
less stringent criteria than usual
for the convergence of the self-consistency 
cycles. These approximations substatially
reduce the computational cost. Therefore, it becomes
possible to relax 
the hundreds of different configurations 
within our family of structures.
We will see below that the energies obtained 
in this first step are reliable enough to select
a set of a few tens of structures containing the 
most promising structural candidates. 
``Accurate" SIESTA calculations 
are performed for these configurations using DZP 
basis sets, a larger k-sampling and well converged self-consistency. 
We can see that the use of a code that utilizes basis sets
of atomic orbitals is instrumental for this gradual increment
of the accuracy of the calculations: while the 
pseudopotentials, density functional, and
basic numerical scheme remains unchanged, the 
size of the basis set (the main factor limiting the size
of the studied systems and the computational time) can 
be varied. This provides a very convenient way to 
deal with the trade-off
between computational speed and accuracy of the calculation.
For the Si(111)-(5$\times$2)-Au we have recently shown
that SIESTA calculations using
DZP basis sets are
in excellent agreement with well 
converged plane-wave calculations.~\cite{riikonen05}
In the present work we confirm this observation 
for the proposed models for the Si(553)-Au surface. 
This seems to confirm that our final structures are
the most favorable within the family considered here.

In the following we present in detail the 
hierarchy of approximations used to
perform our simulations.
As a starting point we 
automatically generate approximate coordinates
for all possible structures
fulfilling certain conditions.  
These conditions will be explained in detail 
in the next section. Our family of structural
models is based
on plausible analogies with the structure of
other related
surface reconstructions like the 
Si(557)-Au or Si(111)-(5$\times$2)-Au.
While reasonable atomic coordinates within the 
plane of the terrace (that we take as the $xy$-plane)
are relatively easy to guess due to the registry 
with the sub-surface bilayer,
the height of the different atoms in the surface bilayer 
is more problematic.  
For this reason, in the first relaxation (named $Sz$ hereafter)
only the atoms in the topmost bilayer are allowed to relax
in the $z$-direction (normal to the terrace). This relaxation
step ensures the interatomic distances  
to be reasonable without changing the topology 
of the surface bilayer and its registry with the underlying 
atoms.  
For the $Sz$ relaxations we use a SZ basis set and a 2$\times$1 
k-sampling.
To further accelarate the simulations the
parameter determining the convergence of 
the density matrix in each relaxation step
(DM.Tolerance~\cite{dmtolerance}) is set to 10$^{-3}$.
The typical value given to this parameter to ensure  
a very good convergence of the self-consistent solution is 10$^{-4}$.
However, we have checked that increasing this value to 10$^{-3}$
only introduces small errors in the calculation of 
energy and forces: for example,
the maximum force difference during the relaxation
of a few representative structures 
of those studied here 
was less than 0.01~eV/\AA\ 
when the two different convergence
criteria were used. In spite of this moderate
effect on the results, in some
cases increasing the value of DM.Tolerance to 
10$^{-3}$ reduces considerably the number of iterations
per self-consistency cycle.
In the next step (named $Sy$ hereafter) all atoms
are allowed to move (except the silicion atoms
in the bottom of the slab and 
the hydrogen atoms directly bonded to them).
However, in order to preserve the topology of the selected
configuration, the positions of the atoms
along the direction parallel to the step edges ($y$-axis) is fixed.  
Other parameters have the same value as in the relaxation $Sz$.
The purpose of the relaxations $Sz$ and $Sy$ is to 
provide a sound initial configuration 
from the coordinates generated automatically.
Using this corrected guess we can proceed 
further allowing all degrees of freedom
in the slab to relax and using a more complete DZ basis set and
accurate 8$\times$4 k-sampling. We call this the $D^{\star}$ relaxation.
Finally, in the $D$ relaxations we further
decrease the tolerance for the covergence
of the elements of the density matrix to its
usual default value in SIESTA~\cite{dmtolerance} of 10$^{-4}$. 
We use a DZP basis set for our most accurate relaxations ($DP$ hereafter).
Adding a polarization shell with $d$ symmetry can 
be especially important to accurately describe ``unusual"
coordinations of the silicon atoms which cannot be described
with simple $sp$ hybridizations.

The use of this series of optimization schemes with ever-increasing
accuracy 
($Sz$$\rightarrow$$Sy$$\rightarrow$$D^\star$$\rightarrow$$D$$\rightarrow$$DP$) 
is much more efficient than starting directly with a relaxation at the $DP$
or similar level. The reason for this efficiency gain is twofold:
{\it i }) the initial coordinates used to start each relaxation
have been optimized at the previous level and, therefore, they 
are an initial guess of increasing quality; 
{\it ii}) the energy estimates obtained with the less accurate
relaxations, already at the $Sy$ level, are accurate enough
to allow discarding many of the possible configurations
in favor of the most favorable models. This is the case even 
if at the $Sy$ level we use
a minimal basis set, a thin slab of two bilayers and  we
fix the atomic coordinate parallel to the step edge in order
to preserve the selected bonding topology.
One has to take into account that a minimal basis for Si contains
only four orbitals, a DZ basis eight orbitals, 
and a DZP basis thirteen orbitals.
Thus the computational cost changes dramatically when changing the 
basis set size. Additional gains are obtained by using smaller
k-samplings and reducing the number of step in each self-consistent cycle.

\subsection{Structural models: a labelling scheme}
\label{sec:labelling}
The structural models that we considered here
for the Si(553)-Au reconstruction are based on
the analogy with other 
similar reconstructions~\cite{crain04} for which
there is more structural information. 
In particular, 
the vicinal Si(557)-Au surface 
reconstruction 
has been extensively studied in recent years,
and its structure
has been established
from X-ray studies~\cite{robinson02}
and DFT calculations.~\cite{sanchezportal02,sanchezportal03,sanchezportal04}
Like in the case 
of the Si(553)-Au surface, the 
steps of the Si(557)-Au run along the $[\bar{1}10]$
and each terrace contains a monatomic chain of gold atoms.
The main difference is the orientation of the 
miscut angle, which is the opposite in both cases.~\cite{crain04}
The size of the terraces is also different, it is larger
in the case of Si(557)-Au allowing for the 
presence of a row of silicon 
adatoms running parallel to the step edge~\cite{robinson02,sanchezportal03}
which seems to be absent in the case of the Si(553)-Au surface.
Therefore, it is clear that both structures can 
be quite different (particularly the step-edge configuration), and
one has to be careful when trying to 
translate structural information from one system to the other.

Fortunately we also have a considerably amount 
of information about the
reconstruction induced by submonolayer deposition of gold 
in flat Si(111). 
The structure
of the 
Si(111)-(5$\times$2)-Au surface, corresponding to a 
larger 
gold coverage than those commented above,
was established experimentally 
by Marks and Plass~\cite{marks95} 
using high resolution electron microscopy
and heavy-atom holography.
Recent first-principles calculations have
shown that the real structure is probably
somewhat different from this original 
proposal~\cite{kang03,erwin03,riikonen05}
and might even depend on the concentration
of silicon adatoms present 
in the surface.~\cite{riikonen05,mcchesney04}
However, there are several common elements
to most of the proposed  models, which are also shared
by the Si(557)-Au reconstruction. The most
important features of the most stable
structural models of these surfaces are: 

{\it i}) the reconstruction only involves the atoms
in the topmost bilayer;

{\it ii}) the gold atoms occupy substitutional positions
in the surface layer, which are much more favorable than 
adatom-like sites;

{\it iii}) positions of gold in the middle of terraces
are favored over step-edge decoration;

{\it iv}) frequent appearance of the
so-called honeycomb chain (HC)~\cite{erwin98} 
structure.

Fig~\ref{fig:fig1}~(a) shows a schematic view 
of an unreconstructed Si(553) surface.
Taking into account the points {\it i})
and {\it ii}) we only
explore here reconstructions generated 
adding an additional bilayer ontop of this 
unreconstructed
substrate. One of the silicon
positions is replaced by a gold atom. Different registries
with underlying bilayer are allowed, as well as, the 
presence of HC structures. Fig~\ref{fig:fig1}~(b), (c)
and (d) show a few possible structures. Structure 
(b) recovers the unreconstructed surface.
Structure (c) presents stacking fault in the middle of
the terrace with the accompanying surface dislocations
with under- and over-coordinated atoms (indicated by 
arrows in Fig.~\ref{fig:fig1}). In panel  
(c) the surface bilayer contains a HC structure in the
middle of the terrace. Notice that the HC reconstruction
also creates a stacking fault towards the $[\bar{1}\bar{1}2]$
direction (i.e., towards the inner part of the terrace).
This stacking fault has to be corrected 
in order to connect with the bulk structure. 
Therefore, it is necessary 
to introduce a surface dislocation with over-coordinated
atoms (marked with an arrow). 

The importance of the HC structure for the gold induced reconstructions
on vicinal Si(111) has been recently emphasized by
Crain {\it et al.}~\cite{crain04}. It is known to form
the step-edge 
of the Si(557)-Au surface~\cite{sanchezportal02,robinson02,sanchezportal03,riikonen07}. 
It is also
a key
ingredient of the most recent proposals for the structure of the
Si(111)-(5$\times$2)-Au reconstruction~\cite{erwin03,riikonen05}.
The HC structure was initially proposed
to explain the low coverage
reconstructions induced by some alkali metals (Li, Na, K), Mg and Ag
on Si(111)~\cite{erwin98}.
We see that the HC structure
involves two unit cells of the unreconstructed Si(111) surface,
with one atom removed from the top Si layer.
This
flattens the surface and removes surface stress.
The two silicon atoms in the center of the HC structure
have a double bond,
which is further stabilized by its 
hybridization with the dangling-bond of atom immediately below.
In the case of reconstructions induced by alkalis, 
the metal atoms donate electrons
to the HC becoming a closed-shell structure and thus 
contributing to the stabilization
of the surface.~\cite{erwin98}
In the case of gold, which has a stronger electron affinity, the situation
is different.
Gold is likely to take electrons away from the silicon structure.
In principle, this does not prevent an
electronic stabilization mechanism:
one electron may be transferred to the 6$s$ Au state, leading
again to a closed-shell structure. However, we are typically 
far
from this {\it ionic} situation. The states
of gold are strongly hybridized with those of
the neighboring silicon atoms creating
several dispersive bands that are,
in principle,
metallic~\cite{sanchezportal02,sanchezportal03,sanchezportal04} and
dominate the photoemission spectra.

In Ref.~\onlinecite{riikonen05_2} we explored
a few structural models for the Si(553)-Au reconstruction
based on an analogy with the Si(557)-Au surface.
Here we want to move a step further and to make 
a comprehensive search among the structural models 
that can be built following the rules ({\it i})-({\it iv}) presented
above. We consider {\it all} possible structures
where the atoms of the topmost bilayer 
present coordinations between 
2 and 4 with other atoms in the same bilayer. 
The final coordination 
depends on the registry with the underlying silicon structure.
One of the
silicon atoms in the unit cell is replaced
by a gold atom. 
The Si(553)-Au reconstruction is known
to suffer several distortions that increase the size
of the unit cell along the step direction as the
temperature is decreased.~\cite{ahn05,snijders06}
However, here we only consider models that preserve the 
$\times$1 periodicity of the silicon substrate along
the steps and, therefore, are relevant to model the high temperature
structure. 

We have developed a simple labelling scheme for 
the family of structural models that fulfill
the criteria presented above.
We can label each structure and 
thus count the total number of 
different trial structures within this family.
Furthermore, this scheme can be easily translated
into a procedure to automatically generate the 
trial geometries.
The basis of our labelling procedure can be found
in Fig.~\ref{fig:fig2}. 
First, the 
possible positions within the {\it xy} plane of the 
surface atoms are discritized
and approximated
by the points of a grid. The grid is formed by nine columns and
two rows. The nine columns correspond to the
positions of
the atoms 
along the $[11\bar{2}]$ direction
in the terrace 
of an unreconstructed 
Si(553) surface (see the Fig.~\ref{fig:fig1}~(b)). 
Second, all possible structures 
created by distributing the 
atoms among the grid points
can be translated into a sequence of nine numbers.
The position along the
horizontal coordinate (column) is
indicated by the order in 
the numerical sequence. The first number corresponds to
the atoms at the step edge.
For a given column, a
``2'' (``4") indicates that a silicon (gold) atom is located
in the higher row, i.e., in the middle of the
rectangular terrace unit cell, 
whereas a``1" (``3") indicates that a silicon (gold)
atom
sits over a grid point in the lower row.
A ``0" indicates that
there are no atoms in that column.
Using this scheme, the unreconstructed
Si(553) surface in Fig.~\ref{fig:fig1}~(b) can
be label as (1,2,2,1,1,2,2,1,1), while structures
in panels (c) and (d) would receive the 
label (1,2,2,2,1,1,2,1,1) and (1,2,2,1,0,1,2,1,1), respectively.
Other examples, 
corresponding to models already studied 
in Ref.~\onlinecite{riikonen05_2}, 
can be found in Fig.~\ref{fig:fig2}.

In principle, using our 
notation we can generate M different
models, with 
\begin{equation}
M= 2^{(N_{Si}+1)}{9! \over {(8-N_{Si})! \: N_{Si}!}}
\end{equation}
and N$_{Si}$ being the number of silicon atoms
in the surface bilayer. Since we always have 
one gold atom, the total number of atoms in the terrace
unit cell is N$_{atm}$=N$_{Si}$+1. 
We consider here structures with N$_{Si}$=7.
In this way, the family of structures 
studied here includes the five models already 
discussed in Ref.~\onlinecite{riikonen05_2}.
Furthermore, having N$_{atm}$=8 is 
a necessary condition to allow for the formation of the 
HC structure, which is one of the common 
building blocks to several gold induced reconstructions
in Si(111) and vicinal Si(111) surfaces (see above).
With N$_{Si}$=7 we have M=18432 different models.
This large number can be considerably reduced 
imposing a few constraints to ensure that the models
represent physically sound structures.
These constraints are:
{\it a}) in order to connect with the bulk structure
the last number of the series must be either
1 or 3;
{\it b}) the dangling-bonds of the underlying silicon bilayer
must be saturated either by an atom or by a dimer as in the 
HC structure, therefore the first number 
of the labelling sequence must be always 1 or 3, 
{\it c}) 
the fifth number must be 0, 1 or 3 (if 0, 
then the neighboring numbers
must be either 1 or 3), and
{\it d})
the third and seventh number must be 0, 2 or 4 (if 0, then
the neighboring numbers in the sequence must be 2 or 4);
{\it e}) to ensure the connectivity within the surface
bilayer, 
a non-zero number in the sequence 
cannot be surrounded by zeros and 
{\it f}) two or more zeros cannot appear
together.
Taken into account conditions ({\it a})-({\it f}), the number 
of possible configurations with N$_{atm}$=8
and N$_{Si}$=7 is reduced to M=210.
In this work we will explore these 210 configurations
using ``fast'' SIESTA calculations, as decribed 
in the previous section, and a few tens of the 
most stable configurations will be selected
to perform more accurate SIESTA and VASP calculations.

Our notation provides information about the connectivity
within the surface bilayer and the registry with the
substrate. From a given sequence of
nine numbers we can generate a trial geometry.
However, we lack information about the heights
of the different atoms. Due to this and
to the discretization of positions 
in the {\it xy}-plane, the bond lenghts and angles
in the automatically generated structures can
considerably depart from the correct values. 
For this reason, as a first step to get 
sound initial configurations we need 
to perform constraint relaxations that, while
preserving the bonding topology of the
selected configuration avoid unphysical bond 
distances and angles. We use the 
$Sz$ and $Sy$ relaxations described in the
previous section for this.

Besides the family of structures described above,
we have explored a few structural models based
on the $\pi$-bonded chain reconstruction of the
Si(111) surface.~\cite{pandey81,himpsel84,tromp84}
In principle, our notation cannot
describe these bonding pattern: it can only 
describe structures which are based on
a ``flat'' surface bilayer. This is partially 
due to the lack of information about the atomic heights.
However, we can modify our notation to describe
the $\pi$-bonded chain structures. This is done allowing for 
a double occupation of the columns and 
is schematically illustrated in Fig.~\ref{fig:fig3}.
These double occupation indicates the position of
the $\pi$-bonded chain in the structure. We still 
have the ambiguity about the relative height of 
atoms in the $\pi$-bonded chain, which is known
to be tilted.
There are two possibilities which are usually 
referred as negative or positive tilt.~\cite{xu04}  
In our notation these two different tilts of
the $\pi$-chain are indicated by the order
of the pair of indices, the second index corresponding to 
the higher atom.
In Fig.~\ref{fig:fig3}
we present a Si(553) surface
reconstructed with the negative tilt chain. 
This negative tilt $\pi$-chain block
corresponds to 
the label
(...,1,0,2,21,1,...), while the label (...,1,0,2,12,1,...) 
denotes the positive tilt structure.
Both configurations are quite similar 
and first-principles calculations predict them
to be almost degenerate in energy and separated
by a very small energy barrier.~\cite{ancilotto90,zitzlsperger97}
Experimentally
the positive tilt structure has been traditionally 
favored.~\cite{himpsel84,tromp84,xu04}
In the case of the Si(553) stepped surface, our
calculations predict the negative tilt structure
to be slightly more stable.

This notation opens
the possibility of generating and studying 
all possible structures
containing the $\pi$-bonded chain. However, we have not pursued
this approach here and we limit 
to consider nine different structural models that are 
obtained after the substitution of one silicon atom by a gold atom
in 
different positions of a 
$\pi$-bonded chain reconstruction 
similar to that shown in Fig~\ref{fig:fig3}. 

\section{Results}

We take advantage of the methodology described in sections 
\ref{sec:relaxations} and \ref{sec:labelling} to make
an extensive search of the structure of 
the Si(553)-Au reconstructions. 
We make a systematic search 
within the 210 structures that can be generated with 
the notation presented in Sec.~\ref{sec:labelling} (models
based on ``flat'' bilayers with different coordinations
and registries with the susbtrate) with seven silicon atoms
and one gold atom in the terrace unit cell.
We also present results from a much more restricted
search for structures based on the $\pi$-bonded chain 
reconstruction.
Finally the most stable structures from these two searches 
are studied using accurate SIESTA and VASP calculations.
We present results for the band structure and the simulated
STM structures of some of the final models.

\subsection{Systematic search: ``flat'' bilayers with N$_{atm}$=8}
\label{sec:systematic}
We first explore the energy of the 210 possible configurations
using our fastest relaxation schemes, $Sz$ and $Sy$,
described in Sec.~\ref{sec:relaxations}. These
calculations transform an initial structure
automatically generated from a given label into 
a physically sound structure. In spite
of the thin slab and minimal basis utilized,
the relative energies obtained at the $Sy$ 
level ($\Delta$E$_1$) already provide
a good guide to eliminate the most unstable
structures. In Table~\ref{tab:tab1} we can 
find a list with the 80 most 
stable configurations ($\Delta$E$_1$ $\leq$33~meV/\AA$^2$)
obtained after $Sy$ relaxations. 
Several of the initial
structures converge to the same configuration,
so these 80 trial structures
give rise to only 40 clearly different structures. 
This is seen in Fig.(\ref{fig:fig5}),
where the plateaus in the energy curve
correspond to this ``lumping'' of several initial
geometries into a single geometry.
This transformation typically takes place by
a displacement of the surface bilayer as a whole, thus 
changing its registry
with the underlying substrate, or by the movement of 
a vacancy
to a neighboring position  
(using our notation this corresponds 
to a transformation (..,0,1,2,..) $\rightarrow$ (..,1,0,2,..)).
This happens for example in the case of the 
(1,2,2,1,1,0,4,1,1) initial 
structure, that transforms into a configuration
that is better described with the label (1,2,0,2,1,1,4,1,1)
and is one of the most stable structures.
This can be seen in Fig.~\ref{fig:fig4} 
(see also Tab.~\ref{tab:tab1}).

The 68 most stable structures, as predicted
using $\Delta$E$_1$, are then calculated again, this time
with more accurate relaxation schemes up to the $DP$ level
(for details, see Tab.~\ref{tab:tab1}).
Hereafter, we will refer to these final relative energies as $\Delta$E$_2$.  
to the $DP$ level. The 
A comparison between $\Delta$E$_1$ and $\Delta$E$_2$
can be found in 
Tab.~\ref{tab:tab1} and Fig.~\ref{fig:fig5}.
Of course, there are important differences between the relative energies
calculated using the minimal basis and constrained relaxations
($Sy$ scheme), 
and those obtained from the fully relaxed structures using 
DZP basis sets ($DP$ scheme). 
However, the overall trend of increasing energy 
is recovered with the more accurate calculations.
We also observe in this case that
several initial geometries give rise 
to the same or very similar final structures. 
However, some of the  approximate degeneracies 
found with the constrained relaxations are removed.

Some of the most stable final geometries are marked with labels
f1-f10 both in Fig.~\ref{fig:fig5} and Tab.~\ref{tab:tab1}. 
These structures are shown in
Fig.~\ref{fig:fig6}.
The fast $Sy$ relaxations ($\Delta$E$_1$)
predicted f1
to be most stable configuration.
It exhibits a HC structure at the step-edge, while
the gold atom is located at a surface dislocation in the middle 
of the terrace. The presence of a surface dislocation is
necessary to recover the bulk stacking disrupted 
by the HC. The position of the gold atom
seems reasonable, gold should be a better
option than silicon to sit at the dislocation 
since gold does not exhibit strong directional 
bonding. However, using
a more complete basis set (already a DZ basis gives the 
correct result) and a thicker slab the f2, f3 and f4 geometries 
become the most favorable structures (they are
almost degenerate and $\sim$4~meV/\AA$^2$ more
stable than f1). 
This points to the 
importance of using more complete (and thus flexible) basis
sets when the coordination of the surface atoms departs
from simple $sp$ hybridization (like in the 
case of the HC structure or at the surface dislocations).

Our results indicate that the configurations 
featuring a HC structure at the step edge (similar to 
the Si(557)-Au structure) are the most stable, at least
within the family of reconstruction considered here.
This confirms the results obtained in our previous  
work~\cite{riikonen05_2}, where we only
studied six different models (one of them was 
the model proposed by Crain {\it et al.}~\cite{crain04}
that exhibits 
a $\times$3 periodicity and, therefore, lies outside 
the present family of models).
The configurations f4 and f2 in the present study 
correspond with the most stable structures obtained 
in Ref.~\onlinecite{riikonen05_2} (named respectively 
I and II in that reference).
Also configurations f8, f9 and f10 correspond to structures III, IV and
V in Ref.~\onlinecite{riikonen05_2}.
Configuration f3, however, is a new configuration 
exhibiting a double honeycomb structure 
similar to that found in some models of 
the Si(111)-(5$\times$2)-Au reconstruction.~\cite{erwin03,riikonen05}

In Sec.~\ref{sec:labelling} we pointed out 
that our labelling scheme excludes, in principle, structures
based on the $\pi$-bonded chain reconstruction. However, in Fig.~\ref{fig:fig6}
we can find one structure where the $\pi$-bonded chain reconstruction has
emerged spontaneously. In configuration f6 the gold
atom is located very close to the step edge. 
The initial
structure 
corresponds to a largely unreconstructed 
terrace. It is well known~\cite{northrup82} that the 
energy barrier for the transformation from the unreconstructed
Si(111) to the $\pi$-bonded chain Si(111)-(2$\times$1) reconstruction is 
very small. Therefore, the appearance
of the $\pi$-bonded chain in this case is not very surprising.

The experimental electronic band structure,
as determined by photoemission 
experiments~\cite{crain03,crain04,ahn05},
presents three bands with parabolic dispersion 
and strong one-dimensional
character. Two of them are similar to those
found for the Si(557)-Au surface and, therefore,
can be assigned to the spin-split bands formed
from the hybridization of Au 6$p$ states with the
$sp$ lobes of the neighboring Si atoms,
as proposed in our recent work.~\cite{sanchezportal04} 
The third band appears centered around
the same point in reciprocal space, but
at higher energies and thus with a lower
occupation around ${1 \over 3}$.

In Fig.~\ref{fig:fig7} we can find the
band structures of the models f2 and f3.
The band structure of the model f4 has been
published elsewhere \cite{riikonen05_2}.
The geometries of f2 and f4 are very similar
(see Fig.~\ref{fig:fig6}) but 
their band structures present 
some small but essential differences.
Both models present 
one dispersive one-dimensional band
coming from the hybridization of gold with its silicon neighbors.
This band can be identified 
with the spin-split bands 
observed in the  Si(553)-Au surfaces~\cite{crain03,barke06} and 
Si(557)-Au~\cite{losio01}.
Two other surface bands appear close to the Fermi level: 
a dispersive band coming from a silicon dangling-bond in the
surface and a band derived from the atoms at the step edge (that has 
a HC structure). These bands are also present for both models.
However, while
in the case of the f4 structure the step-edge band is completely filled
and the dangling-bond band is empty~\cite{riikonen05_2}, 
in the case of the f2 model these two bands cross and the dangling-bond band
has a small occupation of
$\sim\frac{1}{5}$ electrons closer to
the experimental observation.

The population of the different bands can be roughly understood
taking into account the larger electron affinity of
gold and the HC structure~\cite{erwin98} as compared to other
atoms in the surface. 
The population of the dangling-bond band is thus depleted
in favor of the other surface bands.
The differences between the f2 and f4 structures are more
subtle. 
These two models 
only differ in the position of the surface dislocation.
Energetically this structural change has small
consequences and both structures are almost degenerate.
However, it changes the occupation of the dangling-bond band.
In the case of the f2 model the dislocation involves
the gold atoms and 
some of the silicon atoms of the HC structure.
This increases slightly the energy of the bands
associated with the HC and, as a consequence, 
the step-edge band transfers some of its population 
to the dangling-bond band.

Fig.~\ref{fig:fig7} also shows
the band structure of
the f3 model.  
The dangling-bond band is missing in this case. This 
spoils the comparison with experiment. 
The other two bands (gold and step-edge derived)
are very similar to those found for the f2 and f4 models.
This is reasonable taking into account the similar
gold site and structure of the step-egde.

Fig.~\ref{fig:fig8}~(b) shows the band structure
of the f2 model calculated 
including the spin-orbit interaction with the VASP code.
The band structure is in excellent
agreement with that calculated using SIESTA. It also 
confirms that all the bands close to E$_F$ 
with a significant weight in the gold atoms exhibit 
a splitting that has its origin in the spin-orbit
interaction.  

Although the band structure of the model f2 does not exactly
reproduce the photoemission results~\cite{crain03,crain04,ahn05}, 
particularly the characteristic 
band fillings mentioned above, it has some clear
qualitative similaritires with them.
We find two bands with similar dispersions, with their
minima $\sim$1 eV and $\sim$0.5 eV below the Fermi energy 
at the Brillouin-zone boundary.
This corresponds very well to the photoemission data. Furthermore, 
the band with its minima at lower energy shows a 
notable band splitting near the Fermi energy due
to the spin-orbit interaction.
The is in good agreement with the experiments.
However, the band structure of Fig.~\ref{fig:fig8}~(b) has two 
important
differences compared to the photoemission data. 
Firstly, the dangling-bond band presents an important
spin-orbit splitting associated with its appreciable
hybridization with gold (see Fig.~\ref{fig:fig7}).
This splitting is not
observed in the experiments.
Secondly, the theoretical band structure has 
one extra band not seen in 
photoemission. This band is associated with the HC structure
at the step-edge edge and has its 
minimum at $\Gamma$, contrary to the case of the other two bands.

We can also compare the predictions for our models 
with the experimental STM images. 
In Fig.~\ref{fig:fig9} and Fig.~\ref{fig:fig10}
we show the simulated STM
images at different voltages for the
f2 and f4 models. In general we can say that the 
comparison with the available experimental data is satisfactory.
In agreement with the experimental images~\cite{crain06,ahn05,snijders06}
the most prominent feature is the step edge. Within the terrace
we find signals coming from the row of gold atoms
and its neighboring HC structure. The gold chain is seen as
a continuous line for occupied states and presents more structure
for empty states. For both positive and negative voltages we can also distinguish 
a signal coming from the unsaturated silicon dangling bond 
in the terrace close to the step edge. The
step edge and the gold chain could be identified with the 
two parallel chains reported by Snijders {\it et al.}~\cite{snijders06} 
In their recent experiment, 
they observe a strong polarity despendence in the STM images,
and in particular, find zigzag structures
for empty states and ladder configurations for filled states.
In figures \ref{fig:fig9} and \ref{fig:fig10} 
we have sketched some possible candidates
for this kind of behaviour.  
For f2 (Fig.~\ref{fig:fig9}), the ladder structure in filled states 
could result from the registry of the step-edge with respect to the gold atoms.
For empty-states images 
the step-edge becomes slightly more visible and the zigzag geometry
could result from the atoms inside the honeycomb-chain.
In the case of model f4 (Fig.~\ref{fig:fig10}), 
we can identify two entities that change
their registries when going from filled to empty states.  
They are the step-edge and the
complex formed by Au and the surface dislocation.  
They show, similar to the experiment, ladder and  zigzag configurations
respectively for filled and empty states.

\subsection{Restricted search: structures based on the $\pi$-bonded chain }

We first explore two models of the Si(553) stepped 
silicon surface where the terraces are fully covered 
by a (2$\times$1) $\pi$-bonded chain reconstruction. 
They correspond to two 
slightly different arrangements of the $\pi$-bonded chain 
(Fig~\ref{fig:fig11}).
Our  most stable geometry (model p0) correspond to the 
so-called negatively tilted 
$\pi$-bonded chain.~\cite{xu04}
In low energy electron diffraction experiments of the
flat Si(111) surface  
the positive-tilt $\pi$-bonded chain
is usually favored over the negative tilt.~\cite{himpsel84,tromp84,xu04}
However, first-principles DFT calculations predict
both structure to be very close in energy and there are
conflicting claims about
which of them is more stable.~\cite{ancilotto90,zitzlsperger97}

We now proceed to make all possible substitutions
of silicon by gold in the surface bilayer.
This gives rise to nine different models 
for the Si(553)-Au reconstructions that we name
pX, with X=1 corresponding to a substitution 
at the step edge and X$>$1 to a substitution
in the terrace. 
The final energies after accurate
SIESTA relaxations ($DP$ level) 
are listed in Tab.~\ref{tab:tab2}.
In several cases the initial structure was not stable
and suffers strong modifications during the relaxation.
These changes are also summarized in 
Tab.~\ref{tab:tab2}.
Model p4, illustrated in Fig.~\ref{fig:fig12},
is the most stable structure. 
The $\pi$-bonded chain where the gold
substitution takes place 
transforms into a structure  
similar to the  unreconstructed Si(111) surface.
This is reasonable taking into account the 
atomic configuration of gold and confirms 
the tendency of gold to occupy substitutional
positions in the middle of 
the terraces.~\cite{sanchezportal02,sanchezportal03,crain04}

The next most favorable model, p2, 
is illustrated in Fig.~\ref{fig:fig13}.  
Again the $\pi$-bonded chain where the 
substitution took place has disappeared. 
However, this time a configuration reminiscent
of the HC structure has formed in the middle
of the terrace. This transformation is accompanied
by an expansion of the surface bilayer (see the forward
movement of the step-edge atoms in Fig.~\ref{fig:fig13}).

The band structure calculated for the p2 model, 
shown in Fig.~\ref{fig:fig13}~(c),
presents similar characteristics to the experimental
band structure (see the previous section).
We can see in Fig.~\ref{fig:fig13}~(c) that 
the p2 model presents two metallic dispersive bands
centered at point K in the  
zone boundary. One of theses
bands is associated with the Si-Au bonds between
gold and the neighboring silicon atoms
in the step edge.
It has a considerable
gold weight and, therefore, is expected to
exhibit a splitting if the spin-orbit coupling
is taken into account. The other band, however, 
is mainly derived from the unsaturated
dangling bonds of the silicon atoms at the
step edge and presents a smaller filling.
As expected, the 
$\pi$-bonded chain structure that remains in the terrace
does not give rise to any metallic band.

The electron pocket of the step-edge band
around K has an occupation
of $\sim$1/4, quite close to that found in the experiment.
There is another small electron pocket associated with the
step-edge around $\Gamma$ which is not observed
in the experiment. Thus 
we can assign a population of $\sim$0.4 
to the surface bands with a larger weight in the
step-edge atoms.
The dispersive band 
with a mixed silicon-gold character has an occupation of $\sim$0.6
electrons. Therefore, the total population 
associated with the step-edge derived surface bands
is one. This is somewhat surprising if we take into 
account that the Si atoms at the step-edge have, in principle,
three electrons to populate these surface bands.
However, one of these electrons is transferred
to the Au 6$s$ states (which appear several eV below E$_F$) 
and the other electron populates states with large contribution
from the  HC structure (which does not exhibit any metallic band).

Further investigation of the 
p2 model reveals that it is a 
metastable configuration. Performing the 
relaxations with a more stringent force tolerance of 0.01 eV/\AA\ 
results in a more stable structure.
This structure, labelled p2$^\star$ and illustrated in Fig.~\ref{fig:fig14},
exhibits  a strong rebonding of the step edge.
The $\pi$-bonded chain suffers a small translation along
the $[{\bar1}10]$ direction
in order to saturate the dangling bonds at the step edge.
The new position of the atoms of the $\pi$-bonded chain
seems to be a compromise between creating a
surface dislocation and saturating the dangling bonds at the step edge. 
The band structure, shown in Fig.~\ref{fig:fig14}~(c), 
is quite similar to that
of the p2 model. 
The role of the step-edge atoms is now played
by the silicon atoms that formed the
$\pi$-bonded chain. A quite dispersive
surface band, coming from these atoms, appears centered at K.
Another dispersive band with a strong gold weight 
appears at lower energies. Although the topology of 
the band structure resembles that observed in the experiment, 
the filling of the parabolic silicon band is in this
case close 0.4, i.e. larger than that observed
experimentally.
Fig.~\ref{fig:fig14}~(c) also highlights one
of the bands associated with the HC structure,
showing a characteristic dispersion.~\cite{erwin98}

The band structure of model p2$^\star$,
calculated using VASP and including the spin-orbit
interaction (Fig.~\ref{fig:fig8}~(a) )
shows quite good, although not perfect agreement, with 
the experiments. 
As in the case of the f2 model (see above), due to the 
spin-orbit interaction the dispersive bands
suffer a splitting proportional to their weight 
in the gold atoms in the surface. Thus, the splitting 
is much larger for the dispersive band starting at lower energies. 
Again, in contradiction with the experimental observations
we find some degree of splitting also for the 
parabolic band appearing at higher energies, although
this splitting is smaller.
The overall conclusion from Fig.~\ref{fig:fig8}~(a)
is that the p2$^\star$ outperforms
the f2 model in terms of reproducing the photoemission results.

The simulated STM images for this structural model
are shown in Fig.~\ref{fig:fig15}. 
The dependence on the polarity seems to be much larger 
than for models f2 and f4.  The structure
of the STM images becomes more complex when going
from filled to empty states as several ``spots'' appear and
the identification of zigzag or ladder-like
structures becomes quite arbitrary. 

Models p4 and p5 
exhibit a similar rebonding of the step edge as explained
above in the case of p2 and p2$^*$.
However, these new geometries, p4$^\star$ and p5$^\star$, are 
not as stable as p2$^\star$ (see Tab.~\ref{tab:tab2}).
Geometry p5$^\star$ also develops the HC structure 
in the terrace as can be seen in Fig.~\ref{fig:fig16}.

\subsection{Most stable structures: combined SIESTA and VASP results}
In Tab.~\ref{tab:tab3} 
we compare the converged energies of the most stable models found
in the previous sections for the Si(553)-Au 
reconstruction. As discussed in the 
previous sections, these models have been found {\it i}) using
a systematic search among all possible model based on a flat
surface bilayer and eight atoms in the terrace unit-cell and,
{\it ii}) the substitution of gold in different positions of
a $\pi$-bonded reconstruction of the Si(553) terraces.
These two classes of models have different number
of atoms. In order to compare the relative surface energies
we need to define the
chemical potential of silicon. Since the surface should be
in equilibrium with bulk, we have chosen the chemical potential
equal to
the total energy of a silicon atom in bulk.
Since the energy differences are quite small we have decided
to perform the 
calculations of the most stable structure 
with a different methodology in order to cross-check our results.
We have used the plane-wave code 
VASP for this purpose.~\cite{vasp1,vasp2}
We can see that there is an excellent
agreement between SIESTA and VASP
results. This was also observed in 
other similar investigations.~\cite{riikonen05}
Geometries f4 and f2 (see Fig.~\ref{fig:fig6}) are the most stable 
structural models of those found in this extensive
structural search. These two models can be considered degenerate
within the precision of the calculations.

Models f2 and f4 were already obtained as
the most stable ones 
in a much more restricted structural search.~\cite{riikonen05_2}
The present calculations confirm that they are certainly among
the most stable 
reconstructions of Si(553)-Au surface 
that only involve the topmost bilayer.

\section{Conclusions}

We have presented a comprehensive study of the structure
of the Si(553)-Au reconstruction. We have considered
reconstructions restricted
to the topmost bilayer and studied two possible types:
(a) ``flat" surface-bilayer models, where 
the atoms at the topmost bilayer
present different coordinations and different
registries with the underlying bulk, and (b) nine
different models based on substitutions silicon by gold
in different positions of a $\pi$-bonded chain reconstruction
of the Si(553) surface.  

Although the unreconstructed Si(553) surface has
nine inequivalent atomic positions, 
within the familiy of the ``flat" surface-bilayer models,
we have focused on structures containing only eight atoms (seven
silicon and one gold atoms) 
in the topmost bilayer. 
We found
in previous investigations that this
was crucial to reduce the number of dangling bonds
in the final structures and to facilitate the appearance of 
the so-called honeycomb-chain structure~\cite{sanchezportal02}. The 
honeycomb-chain has been proposed 
by us~\cite{sanchezportal02,sanchezportal03,riikonen05,riikonen05_2,riikonen07}
and by other authors~\cite{erwin03,crain04}
as one of the key ingredients of 
the reconstructions induced by gold on Si(111) vicinal 
surfaces and the results presented here seem to confirm this idea.

We have developed a compact
notation that allows us to label and identify all the structures.
This notation is instrumental for the
automatic generation of trial geometries and for counting
the number of inequivalent
structures, i.e., having different bonding topologies.
There are several thousands of flat surface-bilayers models with 
eight inequivalent positions in the topmost bilayer,
however, applying
a few physically motivated constraints we obtained only
210 different models.
All these structures, along with those based on a $\pi$-bonded
reconstruction of the terraces
of the Si(553) surface, 
have been studied using first-principles
density functional calculations with the SIESTA code.
An iterative 
procedure, with a step-by-step increase
of the  accuracy and computational cost of the calculations, was
used to allow for the study of this large number of configurations.

Some conclusions can be drawn from our investigation:

{\it i}) Among all the explored models,
the most stable configurations are those that present a
honeycomb-chain structure at the step edge.
In particular,  models f2 and f4 are the most stable
(see Fig.~\ref{fig:fig6}) with almost degenerate total energies.
The structure of both models is very similar, the sole
difference steming from the relative position of the row of 
gold atoms and the surface dislocation induced by the presence
of the honeycomb-chain.

{\it ii}) Within this group of models, the band structure of model f2 
shows a reasonable agreement with the photoemission 
measurements.~\cite{crain03,ahn05}.  It exhibits two one-dimensional
bands that share some similarities with the experimental results: 
both bands have a parabolic dispersion with
the band minima at the K-point,
one of them having a small fractional occupation
($\sim{1\over5}$).  However, both 
bands present some gold character and thus exhibit an
appreciable spin-orbit splitting. In the experiment
only \emph{one} band shows a clear splitting.

{\it iii})  The STM images 
and their
dependence  
on the bias voltage is satisfactorily 
reproduced by the simulated STM images of  models f2 and f4. 
This is also 
in agreement with the observations of 
Ryang and coworkers~\cite{Ryang07}.
These authors have recently found a good 
agreement between the experimental STM images obtained for
the Si(553)-Au surface at low defect concentration and those
simulated 
for a model similar to our f4 structure.

{\it iv}) Of those models based on a $\pi$-bonded reconstruction of the
Si(553) surface, 
only p2 and p2$^\star$ 
(p2$^\star$ being energetically more favorable)
present a band structure in reasonable 
agreement with the photoemission measurements.
In particular, they exhibit two dispersive one-dimensional
bands similar to the experimental ones: one of them has a filling
close to ${1\over2}$ with a strong gold character, and thus exhibits
a strong
spin-orbit splitting; the other one,
coming from the silicon step edge, has a smaller occupation
and a quite small spin-orbit splitting in good agreement
with the photoemission data.

{\it v}) Simulated STM images of the 
p2$^\star$ model present a strong polarity dependence.
Unfortunately, patterns appearing are hard to identify with those
seen in the experiments.

In summary, it seems that model f2 is 
a good candidate for the structure of the 
Si(553)-Au surface. 
This model presents
the lowest total energy and its band structure
and simulated STM images are in general agreement with 
the existing experimental 
data. The agreement, however, is not perfect.  
For example, model p2$^*$ seems to provide a better comparison with the 
photoemission data.  From a more general perspective, our data clearly
point to the importance of the honeycomb-chain 
structure to stabilize the surface and, probably, to the 
presence of a complex rebonding of the step edge,
like that found in our pX$^\star$ models. If this the 
case, the structure of the step edge could appreaciably
depart from that of the initial models considered 
in the present study. 
More theoretical and
experimental work is therefore necessary to determine if model
f2 indeed corresponds to the high temperature structure of 
the Si(553)-Au surface reconstruction.

%Tables
\begin{table*}\small
\begin{tabular}{ll}

\begin{tabular}{|l|l|l|l|}
\hline
No. & Initial  & $\Delta$E$_1$ 
& $\Delta$E$_2$ \\
 & configuration & (meV/\AA$^2$) & (meV/\AA$^2$) \\
\hline\hline
1 & 1,2,2,0,1,1,4,1,1 & 0.00 & 0.00 \\
2 & 1,2,2,1,0,1,4,1,1 & 0.00 & 0.00 \\
3 & 1,2,2,1,1,0,4,1,1 & 0.00 & 0.00 \\
4 & 1,2,0,2,1,1,4,1,1 & 0.00 & 0.00 ~~ (f1)\\
5 & 1,0,2,2,1,1,4,1,1 & 0.01 & -0.01 \\
6 & 1,2,2,0,1,3,2,1,1 & 5.51 & -4.27 \\
7 & 1,2,2,1,0,3,2,1,1 & 5.51 & -4.29 \\
8 & 1,0,2,2,1,3,2,1,1 & 5.51 & -4.26 \\
9 & 1,2,2,1,3,0,2,1,1 & 5.53 & -4.30 \\
10 & 1,2,0,2,1,3,2,1,1 & 5.54 & -4.28 ~ (f2)\\
11 & 1,2,2,1,1,0,2,3,1 & 6.08 & -4.02 \\
12 & 1,2,0,2,1,1,2,3,1 & 6.09 & -4.01 ~ (f3) \\
13 & 1,2,2,0,1,1,2,3,1 & 6.09 & -4.02 \\
14 & 1,2,2,1,0,1,2,3,1 & 6.09 & -4.01 \\
15 & 1,0,2,2,1,1,2,3,1 & 6.12 & -4.02 \\
16 & 1,0,2,2,1,4,2,1,1 & 6.21 & -4.42 \\
17 & 1,2,0,2,1,4,2,1,1 & 6.23 & -4.46 ~ (f4) \\
18 & 1,2,2,0,1,4,2,1,1 & 6.29 & -4.45 \\
19 & 1,2,0,2,3,2,2,1,1 & 7.47 & 1.75 ~~ (f5) \\
20 & 1,2,2,0,3,2,2,1,1 & 7.49 & 1.73 \\
21 & 1,0,2,2,3,2,2,1,1 & 7.49 & 1.74 \\
22 & 1,0,4,1,1,2,2,1,1 & 14.69 & 9.55 ~~ (f6) \\
23 & 1,2,2,1,1,0,2,4,1 & 14.76 & -0.20 \\
24 & 1,2,2,1,0,1,2,4,1 & 14.76 & -0.20 \\
25 & 1,2,2,1,1,2,0,4,1 & 14.77 & -0.21 \\
26 & 1,2,0,2,1,1,2,4,1 & 14.77 & -0.17 ~ (f7) \\
27 & 1,2,2,0,1,1,2,4,1 & 14.78 & -0.20 \\
28 & 1,2,2,1,1,2,4,0,1 & 14.82 & -0.18 \\
29 & 1,0,2,2,1,1,2,4,1 & 15.41 & -0.20 \\
30 & 1,0,2,2,1,3,2,2,1 & 17.49 & 16.12 \\
31 & 1,2,0,2,1,3,2,2,1 & 17.49 & 16.13 \\
32 & 1,2,2,0,1,3,2,2,1 & 17.50 & 5.77 \\
33 & 1,0,2,3,1,2,2,1,1 & 21.98 & 1.89 \\
34 & 1,0,2,4,1,2,2,1,1 & 22.57 & 5.00 \\
35 & 1,2,2,1,1,2,0,2,3 & 23.49 & 12.80 \\
36 & 1,2,2,1,1,2,2,0,3 & 23.49 & 12.79 \\
37 & 1,2,2,1,0,1,2,2,3 & 23.64 & 11.70 \\
38 & 1,2,2,0,1,1,2,2,3 & 23.68 & 11.55 \\
39 & 1,2,2,1,1,0,2,2,3 & 23.72 & 12.78 \\
40 & 1,0,2,2,1,1,2,2,3 & 23.79 & 12.20 \\
\hline
\end{tabular}

&

\begin{tabular}{|l|p{2.3cm}|l|l|}
\hline
No. & Initial & $\Delta$E$_1$ 
& $\Delta$E$_2$  \\
& configuration & (meV/\AA$^2$) & (meV/\AA$^2$)  \\
\hline\hline
41 & 1,2,2,0,3,1,2,1,1 & 23.88 & 7.47 \\
42 & 1,2,2,3,0,1,2,1,1 & 23.90 & 7.48 \\
43 & 1,2,2,3,1,0,2,1,1 & 23.91 & 7.47 \\
44 & 1,2,0,2,3,1,2,1,1 & 23.91 & 7.47 \\
45 & 1,0,2,2,3,1,2,1,1 & 23.91 & 7.47 \\
46 & 1,2,0,2,1,1,2,2,3 & 24.13 & 12.18 \\
47 & 1,2,0,2,1,2,4,1,1 & 25.23 & 7.76 \\
48 & 1,0,2,2,1,2,4,1,1 & 25.24 & 7.77 \\
49 & 1,2,2,0,1,2,4,1,1 & 25.24 & 7.77 \\
50 & 1,2,0,2,1,2,2,3,1 & 25.41 & 4.17 ~ (f8)\\
51 & 1,2,2,0,1,2,2,3,1 & 25.41 & 4.13 \\
52 & 1,0,2,2,1,2,2,3,1 & 25.42 & 4.13 \\
53 & 1,2,0,4,1,2,2,1,1 & 25.67 & 11.56 \\
54 & 1,2,4,0,1,2,2,1,1 & 25.72 & 11.50 \\
55 & 1,4,2,0,1,1,2,1,1 & 26.31 & 9.40 \\
56 & 1,4,2,1,0,1,2,1,1 & 26.34 & 9.40 \\
57 & 1,4,2,1,1,0,2,1,1 & 26.36 & 9.40 ~ (f9) \\
58 & 1,4,0,2,1,2,2,1,1 & 27.35 & 12.50 \\
59 & 1,0,4,2,1,2,2,1,1 & 27.35 & 12.49 \\
60 & 1,2,4,0,1,1,2,1,1 & 27.37 & 15.34 \\
61 & 1,4,2,0,1,2,2,1,1 & 27.40 & 12.50 \\
62 & 1,2,2,3,1,2,2,0,1 & 27.62 & 2.66 \\
63 & 1,2,2,3,1,2,0,2,1 & 27.63 & 2.66 ~ (f10) \\
64 & 1,2,2,3,0,1,2,2,1 & 27.63 & 2.68 \\
65 & 1,4,2,1,1,0,2,2,1 & 28.08 & 11.52 \\
66 & 1,0,4,2,1,1,2,2,1 & 28.09 & 9.70 \\
67 & 1,4,2,1,0,1,2,2,1 & 28.09 & 9.75 \\
68 & 1,4,2,0,1,1,2,2,1 & 28.09 & 9.75 \\
69 & 1,2,2,1,0,3,2,2,1 & 28.91 &  \\
70 & 1,2,0,4,1,1,2,1,1 & 29.48 &  \\
71 & 1,2,4,1,0,1,2,1,1 & 29.48 &  \\
72 & 1,2,4,1,1,0,2,1,1 & 29.51 &  \\
73 & 1,0,2,1,1,4,2,1,1 & 32.35 &  \\
74 & 1,1,2,4,1,0,2,2,1 & 33.05 &  \\
75 & 1,1,2,4,1,2,2,0,1 & 33.05 &  \\
76 & 1,1,2,4,1,2,0,2,1 & 33.06 &  \\
77 & 1,2,0,2,3,1,2,2,1 & 33.43 &  \\
78 & 1,2,2,0,3,1,2,2,1 & 33.43 &  \\
79 & 1,0,2,2,3,1,2,2,1 & 33.43 &  \\
80 & 1,2,0,2,1,2,2,1,3 & 33.48 &  \\
\hline
\end{tabular}

\\

\end{tabular}
\caption{\label{tab:tab1}
Results from the automatic structural search.
Total energies $\Delta E_1$ are obtained using a thin slab of 
only two silicon bilayers and the fastest (and less accurate)
relaxations ($Sz$ and $Sy$). 
Only the 80 most stable configurations (out of the
total 210 studied structures) are included in this table, with
a maximum energy difference of $\Delta E_1$ $\sim$~33~meV/\AA$^2$.
The configurations are numbered according to their
predicted stability. The initial configurations are
labelled using the notation developed in Sec.~\ref{sec:labelling}.
The 68 most stable structures according to $\Delta E_1$ are also calculated
using a thicker slab of four bilayers and our 
accurate SIESTA calculations 
($D^\star$$\rightarrow$$D$$\rightarrow$$DP$), 
resulting in the $\Delta$E$_2$ energy differences.
Several initial configurations converge to a single
structure. 
Some of the final structures are indicated with labels f1-f10
(see also figures \ref{fig:fig5} and \ref{fig:fig6}).
}
\end{table*}

\begin{table*}
\begin{tabular}{|l|l|l|l|l|}
\hline
Name & Configuration & $\Delta$E (meV/\AA$^2$) \\
\hline\hline
p & (1,0,2,12,1,0,2,12,1) & 1.82	\\
p0 & (1,0,2,21,1,0,2,21,1) & 0.00	\\
\hline\hline
p1 & (3,0,2,21,1,0,2,21,1) &14.92  \\	
p2 & (1,0,4,21,1,0,2,21,1)$\rightarrow$(1,4,2,1,1,0,2,21,1) &4.13 \\
p3 & (1,0,2,41,1,0,2,21,1)$\rightarrow$(1,2,4,1,1,0,2,21,1) &8.15 \\
p4 & (1,0,2,23,1,0,2,21,1)$\rightarrow$(1,2,2,3,1,0,2,21,1) &0.00 \\
p5 & (1,0,2,21,3,0,2,21,1)$\rightarrow$(1,2,0,2,1,3,2,21,1) &7.07 \\
p6 & (1,0,2,21,1,0,4,21,1)$\rightarrow$(1,2,2,1,1,4,2,1,1)  &9.13 \\
p7 & (1,0,2,21,1,0,2,41,1) &8.30 \\
p8 & (1,0,2,21,1,0,2,23,1) &11.98 \\
p9 &(1,0,2,21,1,0,2,21,3)  &15.90 \\
\hline
\hline
p2$^\star$ & &-4.28	 \\
p4$^\star$ & &-2.69	\\
p5$^\star$ & &0.38	\\
\hline
\end{tabular}
\caption{
Relative surface energies 
of different structures based on 
a $\pi$-bonded chain reconstruction of the terraces of 
the Si(553) and Si(553)-Au surfaces. 
Models p and p0 correspond to the clean silicon surface.
pX corresponds to the substitution of a gold atom
in position ``X" of structure p0. During the relaxation
process several of these structures transform into 
configurations with a different bonding topology.
This change is also indicated.
pX$^\star$ refers to pX configurations after the
formation of bonds between the step edge and 
the neighboring $\pi$-chain structure.
All energies correspond to our most accurate
SIESTA calculations.
\label{tab:tab2}}
\end{table*}

\begin{table}
\begin{tabular}{|l|l|l|}
\hline
Name & \multicolumn{2}{c|}{ $\Delta$E (meV/\AA$^2$) }\\
\hline 
& SIESTA & VASP \\
\hline\hline
p2*		&4.85		&4.93	\\	
p4*		&6.44		&6.54	\\
p5*		&9.51		&9.13	\\
f1		&4.63		&4.27   \\
f2		&0.17		&-0.08  \\
f3		&0.42		& 0.51	\\
f4		&0.00		&0.00	\\
\hline
\end{tabular}
\caption{
Relative surface
energies of our most stable models calculated using both SIESTA and VASP.
\label{tab:tab3}}
\end{table}

\clearpage
%Figures
\begin{figure}
\caption{
(a) Schematic view of the unreconstructed 
Si(553) surface. The steps run parallel to the $[{\bar1}10]$ 
direction and are oriented towards
the $[11{\bar2}]$ direction. The rectangle indicates
the ``unit cell"  {\it within} the terrace, containing nine
inequivalente silicon sites (with four unsaturated dangling-bonds).
Panels (b), (c) and (d) show configurations
generated by adding a silicon bilayer on top of the structure 
in panel (a). The different reconstructions 
explored in this paper are obtained by
changing the structure of this surface bilayer 
and/or its registry with the underlying atoms.
Open triangles represent the topmost
atoms and solid circles the higher atoms in the second bilayer.
These structures (and all the structures considered
in the present work) preserve a
$\times$1 periodicity along the step.
Structure (b) recovers an unreconstructed silicon structure.
Structure (c) presents a surface dislocation 
close to the step egde that 
generates a stacking fault (SF) that is
later
corrected creating another surface dislocation in order
to connect with the bulk structure. Surface dislocations
create under- and over-coordinated atoms
which appear indicated by arrows. 
Model (d) presents a honeycomb-chain (HC) structure and the 
accompanying surface dislocation.
\label{fig:fig1}}
\end{figure}

\begin{figure}
\caption{
(a) and (c) show  two possible structures
for the Si(553)-Au surface already explored in
Ref.~\onlinecite{riikonen05_2}. These structures are 
characterized by the presence of a HC structure
close to the step-edge.
Solid lines indicate the bonds between atoms in 
the topmost bilayer of a given terrace, thin dashed lines
correspond to the underlying silicon bilayer, and
thick dashed lines indicate a few bonds of 
the upper terrace.
The small solid circles mark the positions of the higher 
silicon atoms in the underlying bilayer.
The large open circles mark the substitutional sites
occupied by the gold atoms. The rectangles 
mark the terrace ``unit cell".
Panels (b) and (d) schematically explain how the structure
of the surface bilayer can be translated
into a sequence of nine numbers. First the 
possible positions of the atoms are approximated
by the points of a grid. The grid is formed by nine columns and 
two rows. The position along the 
horizontal coordinate is
indicated by the order in
the numerical sequence. The first number corresponds to 
the atoms at the step edge.
A ``2'' (``4") indicates that a silicon (gold) atom is located
in the higher row, i.e., in the middle of the 
rectangular cell. A ``1" (``3") indicates that a silicon (gold)
atom
sits over a grid point in the lower row.
A ``0" indicates that
there are no atoms in that column. 
\label{fig:fig2}}
\end{figure}

\begin{figure}
\caption{
(a) Scheme of a $\pi$-bonded chain reconstruction of the  Si(553) surface,
and (b) the proposed notation for such structure. The 
presence of the $\pi$-bonded chain at a certain location
is indicated by the
double occupation of the corresponding column,
the second atom of this pair occupies
the higher position along the $\pi$-chain.
\label{fig:fig3}}
\end{figure}

\begin{figure}
\caption{
Relaxation of the structure generated
from the label (1,2,2,1,1,0,4,1,1) at the $Sy$ level.  
Insets (a) and (b) present a top and a lateral view
of the automatically generated
structure.  Inset (c) presents the structure after the 
relative heights of the atoms have been corrected by the 
$Sz$ relaxation. 
In point (d) we have already reached a structure
similar to the configuration that we want to explore.
However, this configuration is not stable and transforms
by the displacement of the surface bilayer (inset (e)) into a different
structure. The final structure, shown in inset (f), is better described by the  
(1,2,0,2,1,1,4,1,1) label. 
\label{fig:fig4}
}
\end{figure}

\begin{figure}
\caption{
Relative energies after fast $Sy$ ($\Delta$E$_1$, solid circles) and 
accurate $DP$ ($\Delta$E$_2$, open diamonds) relaxations
for the systems listed in Tab.~\ref{tab:tab1}.
The system labels f1-f10 correspond to those of Tab.~\ref{tab:tab1}
and Fig.~\ref{fig:fig6}.
\label{fig:fig5}}
\end{figure}

\begin{figure}
\caption{
Final geometries (at the $DP$ level) for a few
selected configurations from those listed in Tab.~\ref{tab:tab1}
and Fig.~\ref{fig:fig5}.
In the case of the f2 and f3 structures some symbols are assigned to 
a few atoms in the surface. They will be used  
to indicate the main character of the different bands
(see Fig.~\ref{fig:fig7}).
\label{fig:fig6}}
\end{figure}

\begin{figure}
\caption{
Band structures of models f2 and f3 calculated using SIESTA.
The atomic character of different bands is indicated using the 
symbols in Fig.~\ref{fig:fig6}. Filled squares indicate the contribution
coming from the gold atoms and their silicon neighbors, 
open circles that coming
from the atoms at the step edge,
and open diamonds that of some silicon atoms in the surface presenting 
unsaturated bonds.
\label{fig:fig7}}
\end{figure}

\begin{figure}
\caption{
Band structures calculated using VASP and including the
spin-orbit coupling for (a) model p2$^\star$ and (b) model f2.
\label{fig:fig8}}
\end{figure}

\begin{figure}
\caption{
Simulated STM images, calculated using SIESTA, for model f2.
The bias voltage is indicated in Volts in each panel.
Some features having
a strong polarity dependence
are indicated by filled circles.
\label{fig:fig9}}
\end{figure}

\begin{figure}
\caption{
Simulated STM images, calculated using SIESTA, for model f4.
The bias voltage is indicated in Volts in each panel.
Some features having
a strong polarity dependence
are indicated by filled circles.
\label{fig:fig10}}
\end{figure}

\begin{figure}
\caption{
Two possible models 
of the Si(553) surface 
based on a $\pi$-bonded chain reconstruction of
the terraces. Model p in panel (a) and 
model p0 in panel (b).
\label{fig:fig11}}
\end{figure}

\begin{figure}
\caption{
Top (a) and side (b) view of the relaxed structure of the p4 model.
\label{fig:fig12}}
\end{figure}

\begin{figure}
\caption{
Top (a) and side (b) view of the relaxed structure of the p2 model.
(c) Electronic band structure. The symbols highlight
those surface bands with an appreciable weight from the 
surface atoms marked with the same symbols in panel (b).
The inset shows a schematic representation of the
most prominent surface bands.
The HC structure in the middle of the terrace is
indicated by a box. 
\label{fig:fig13}}
\end{figure}

\begin{figure}
\caption{
Top (a) and side (b) view of the relaxed structure 
of the p2$^\star$ model, 
obtained after the reconstruction of the step edge.
(c) Electronic band structure. The symbols indicate
those surface bands with an appreciable weight from the
surface atoms marked with the same symbols in panel (b).
The inset shows a schematic representation of the
most prominent surface bands.
The HC structure is highlighted by a box. 
\label{fig:fig14}}
\end{figure}

\begin{figure}
\caption{
Simulated STM images, calculated using SIESTA, for model p2$^\star$.
The bias voltage is indicated in Volts in each panel.
\label{fig:fig15}}
\end{figure}

\begin{figure}
\caption{
Side view of the  p5$^*$ model. 
The HC structure emerges 
and is highlighted by a box.
\label{fig:fig16}}
\end{figure}

\end{document}